
%
\input phyzzx
\pubnum{OU-HET 169 \cr gr-qc/9210003}
\date={September, 1992}
\titlepage
\vskip0.5cm
\title{The Wave Function of the Universe by the New Euclidean
Path-integral Approach in Quantum Cosmology}
\vskip1.0cm
\author{Atushi Ishikawa$^{1}$,\ and \ Haruhiko Ueda$^{2,3}$}

\vskip1.0cm

\centerline {\it $^1$ Department of Physics}
\centerline {\it Osaka University, Toyonaka, Osaka 560, Japan}

\vskip0.5cm

\centerline {\it  $^2$ Uji Research Center, Yukawa Institute for
Theoretical Physics}
\centerline {\it Kyoto University, Uji 611, Japan}

\vskip0.5cm

\centerline {\it $^3$ Department of Physics}
\centerline {\it Hiroshima University, Higashihiroshima 724, Japan }

\vskip1.5cm

\abstract{The wave function of the universe is evaluated by using
the Euclidean path integral approach.
As is well known, the real Euclidean path integral diverges
because the Einstein-Hilbert action is not positive definite.
In order to obtain a finite wave function, we propose a new
regularization method and calculate the wave function of the Friedmann-
Robertson-Walker type minisuperspace model.
We then consider a homogeneous but anisotropic type minisuperspace model,
which is known as the Bianch type I model.
The physical meaning of the wave function by this new regularization
method is also examined.}

\medskip

\vfill\eject

\REF\Revone{S.W.~Hawking, in {\it 300 Years of Gravitation}, edited by
S.W.~Hawking and W.Isreal (Cambridge, University Press, Cambridge;
1987);}
\REF\Revtwo{J.J.~Halliwell, in {\it proceedings of the Jerusalem
Winter School on Quantum Cosmology and Baby universe},
edited by T.Piran (World Scientific, Singapore; 1990);}
\REF\Hawwd{J.B.~Hartle and S.W.~Hawking, {\it Phys.~Rev.} {\bf D~28}
(1983) 2960;}
\REF\Hawtwo{S.W.~Hawking, {\it Nucl.~Phys.} {\bf B239}
(1985) 15;}
\REF\Gib{G.W.~Gibbons, S.W.~Hawking and M.J.~Perry, {\it Nucl.~Phys.}
{\bf B~138} (1978) 141;}
\REF\Hor{G.T. Horowitz, {\it Phys.~Rev.} {\bf D~31} (1985) 1169;}
\REF\HAL{J.J.~Halliwell, {\it Phys.~Rev.} {\bf D~38} (1988) 2468;}
\REF\HLone{J.J.~Halliwell and J.~Louko, {\it Phys.~Rev.} {\bf D~39}
(1989) 2206;}
\REF\HLtwo{J.J.~Halliwell and J.~Louko, {\it Phys.~Rev.} {\bf D~40}
(1989) 1868;}
\REF\HLthr{J.J.~Halliwell and J.~Louko, {\it Phys.~Rev.} {\bf D~42}
(1990) 3997;}
\REF\Gibtwo{G.W.~Gibbons and J.B.~Hartle, {\it Phys.~Rev.}
{\bf D~42} (1990) 2458;}
\REF\Hawone{S.W.~Hawking, {\it Phys.~Lett.} {\bf 151~B}
(1983) 2960;}
\REF\Ser{M.~Seriu, Kyoto University preprint KUNS-1077 (1991);}

\sequentialequations
\def\int{\intop\nolimits}

Quantum cosmology is one of the most fascinated subjects
in modern physics. \refmark {\Revone, \Revtwo}
In quantum theory, the wave function characterizes each theory.
Therefore the main purpose in quantum cosmology is to evaluate the
wave function of the universe. As is well known, the
wave function of the universe is obtained as the solution
of the Wheeler-DeWitt second-order functional differential equation.
This equation, however, is very difficult to solve even in minisuperspace
models, although in minisuperspace models the Wheeler-DeWitt equation
is reduced to an ordinary differential equation.
In addition, we have to
know the initial condition of the universe which is hard to understand
completely. Another approach, which gives the wave function by
the Euclidean path integral, has been proposed by Hartle and Hawking
\refmark {\Hawwd, \Hawtwo}.
They claimed that
the path integral should be performed over regular and closed four-dimensional
manifolds.
This idea is called the boundary condition of no-boundary.
The four-dimensional manifold
must be Euclidean to avoid the cone singularity at
an initial point.
Their proposal solves the initial condition problem
but the real Euclidean path integral does not converge because the
Einstein-Hilbert action is not positive definite.
The question is therefore whether we can construct the Euclidean path
integral approach which is free from a divergence problem.

A first way to avoid this difficulty was proposed by Gibbons, Hawking
and Perry \refmark {\Gib}. They suggested that the path integral
could converge
if we performed the conformal rotation. This proposal, however, has
several objections \refmark {\Hor}.
Hartle proposed another path integral approach. He suggested that the path
integral should be taken along the steepest-descent path in the space
of complex four-metrics.
Following the Hartle's alternative proposal, we can make
the Euclidean path integral convergent by changing an integration contour
from a real axis to a curve on a complex plane.
Halliwell and Louko
calculated the wave function of the universe
following this method \refmark {\HLone, \HLtwo, \HLthr}.
The integral along this
complex curve is so difficult to
perform that they evaluated the wave function in a saddle point
approximation \refmark {\HLone}.
Moreover the physical interpretation of this wave function
is hard because the path integral is defined on the complex
four-manifold \refmark {\Gibtwo}.

The contour changing method seems not to complete
and it is worthy
investigating a possibility of another regularization method.

In this paper, we propose a new regularization method and evaluate
the wave function in minisuperspace models.
We first propose
a new regularization method and calculate the wave function of the
Friedmann-Robertson-Walker type minisuperspace model.
We then consider a more general case, {\it i.e.} a homogeneous
but anisotropic type minisuperspace model, which is known as the Bianch type
I model.
We give a physical interpretation of the
wave function by this new reguralization method.

We first calculate the wave function of the closed
Friedmann-Robertson-Walker type minisuperspace model \refmark {\HAL,
\HLone}. The metric of the Friedmann-Robertson-Walker type is
$$
ds^2={\sigma}^2 \Bigl \{ {N^2(\tau) \over q(\tau)} d \tau^2
            +q(\tau) d \Omega_3^2  \Bigr \},
\eqn\ee
$$
where ${\sigma}^2=2G/3{\pi}$ and $d \Omega_3^2$ is the metric
on the unit three-sphere.
The Einstein-Hilbert action with the rescaled cosmological constant $\lambda$
then becomes
$$
I[q(\tau)]={1\over 2}\int_{\tau'}^{\tau''}d \tau N
     \Bigl [ -{\dot q^2 \over 4N^2} +\lambda q-1 \Bigr ].
\eqn\ee
$$
We can rewrite this action to the Hamiltonian form as follows.
$$
I[p(T),q(T)]=\int_{0}^{T} d \tau \Bigl \{p
\dot q -{1\over 2} (-4 p^2- \lambda q +1 ) \Bigr \},
\eqn\ee
$$
where $T=N \cdot (\tau''-\tau')$ and $p=- \dot q/4N $.
The Wheeler-DeWitt equation can be easily obtained by replacing $p$
in the Hamiltonian with $-d/dq$, therefore the Wheeler-DeWitt equation
in the Friedmann-Robertson-Walker type becomes
$$
H \Psi ={1 \over 2} \Bigl [ -4 {d^2 \over dq^2} -\lambda q
         +1 \Bigr ] \Psi =0.
\eqn\ee
$$
The wave function of the universe can be obtained if one solves this
equation \refmark {\Hawone}.
As is stated before, we can estimate the wave
function by means of the path integral
$$
\Psi = \int dT \int Dp Dq e^{-I[p,q]}
     \equiv \int dT \psi(T) ,
\eqn\ee
$$
instead of solving the Wheeler-DeWitt equation \refmark {\HAL}.
The Wheeler-DeWitt equation will be a guide for the path integral
to define its contour.
As is easily
seen, the integration with respect to $T$ is not easy to evaluate.
We first consider the $p$ and $q$ integration \refmark {\Ser}.
To perform the $p$ and $q$ integration, we divide the
interval $T$ into $L+1$ pieces. Then $\psi(T)$ is
$$\eqalign{
\psi(T)=\lim_{L \rightarrow \infty}&
        \int \prod_{i=1}^L dq_i \prod_{i=0}^L
        {dp_i \over 2 \pi }                 \cr
       &\exp \Bigl \{- \triangle T
        \sum_{i=0}^L (p_i {q_{i+1}-q_i \over \triangle T} +
        2 p_i^2+{\lambda \over 2} q_i -{1 \over 2}) \Bigr \}, \cr
}\eqn\ee
$$
where $\triangle T=T/(L+1),~T_i=i \cdot \triangle T,~
q_i=q(T_i),~p_i=p(T_i)$.
The $p$ integration is carried out immediately because of the Gaussian form
with respect to $p$.
In order to perform the $q$ integration, we use a following formula which
is obtained by the mathematical induction
$$\eqalign{
\int dx_1&  dx_2 \cdots dx_n
          e^{-\alpha \{ (x_1-a)^2+\beta a
           +(x_2-x_1) +\beta x_1 + \cdots
           +(b-x_n)^2+\beta x_n \} } \cr
&={1 \over (n+1)^{1/2}} \bigl ( {\pi \over \alpha} \bigr )^{n/2}
   \times \cr
&~~~  \exp \Bigl \{ {-\alpha \over n+1} (b-a)^2
  -\alpha \beta \bigl [ {n+2 \over 2} a +{n \over 2} b
  -{n \over 48} (n+1)(n+2) \beta \bigr ] \Bigr \}.\cr
}\eqn\ee
$$
We can evaluate the $p,q$ integration completely and the result is
$$
\psi (T)=\Bigl \{ {1 \over 8 \pi T} \Bigr \}^{1/2} e^{-I(T)},
\eqn\ee
$$
where
$$
I(T)= {\lambda^2 \over 24} T^3 + \bigl \{
         {\lambda (q''+q') \over 4 } - {1 \over 2} \bigr \} T
         -{ (q''-q')^2 \over 8T }.
\eqn\ee
$$
In Ref.[8], Halliwell and Louko divided the variables $p(T)$
and $q(T)$ into the classical part and the quantum one.
On the other hand, we can calculate the $p,q$ integration directly
by using above induction (7).

We now evaluate the $T$ integration.
$T$ is the rescaled lapse function and the range of the $T$
integration must preserve
the time reparametrization invariance.
The $T$ integration should be performed from $-\infty$ to $+\infty$ naively.
This integration, however,
diverges if we take the contour along the real axis.
{}From this reason, a regularization is necessary to
remove this divergence.
In Ref.[\HLone, \HLtwo, \HLthr],
the integration is regularized
by means of changing a contour of the $T$ integration
and, by using a saddle point approximation, they evaluated the wave function.
While the contour changing method is a very natural regularization,
the full integration is very difficult.
It seems to be hard to study the wave function beyond
a saddle point approximation.
We adopt another strategy, {\it i.e.}
instead of changing the contour directly,
we extend the integrated function $ \psi(T) $
in (8) such that the $T$ integration will converge.

We introduce a complex parameter $\alpha$ and define the wave function
as an analytic function with respect to $\alpha$ as follows
(up to a normalization factor).
$$\
\Psi (\alpha) = \int_{-\infty}^{\infty} { dT \over {(\alpha T)^{1/2}} }
                  e ^{-I(\alpha T)},
\eqn\ee
$$
where the rescaled lapse function $T$ remains real and
the complex parameter $\alpha$ must be determined
in such a way that the $T$ integration
will converge. The strategy is as follows.
We temporally fix $\alpha$ as a constant complex
number during the integration with respect to $T$.
After the integration, we take the analytic continuation
$ \alpha \rightarrow 1 $ and get the wave function.

Let us perform the above strategy and obtain the wave function of the
Friedmann-Robertson-Walker minisuperspace.
In this case, unfortunately, we cannot carry out the integration completely.
However we can use the Wheeler-DeWitt equation and
obtain the wave function, which is equivalent
to calculating the integration.
This method was proposed by Halliwell and Louko \refmark {\HLone}.
Apparently the wave function is symmetric with respect
to $q'=q(\tau')$ and $q''=q(\tau'')$.
In addition, the wave function satisfies the Wheeler-DeWitt equation
at $q'$ and $q''$.
By rescaling $T \rightarrow \lambda^{-2/3} 2^{1/3} T$
and introducing a new variable
$z \equiv (1- \lambda q)/(2 \lambda)^{2/3} $,
the wave function as a solution of the Wheeler-DeWitt equation
will be expressed as Airy function's products,
$$\eqalign{
\Psi = &a Ai(z'') Ai(z') +b Bi(z'') Bi(z')                 \cr
       &~~~~~+c \bigl [ Ai(z'') Bi(z') +Bi(z'') Bi(z') \bigr ].
}\eqn\ee
$$
In order to get coefficients $(a,b,c)$, we take $z'=z''=z$ and
expand (11) with $z$ around $z=0$. On the other hand, (10) will be
$$
\Psi ( \alpha ) = {1 \over \alpha} \int_{- \alpha \infty}^{\alpha \infty}
     {dT \over T^{1/2}} \exp \Bigl \{-
     {1 \over 12} T^3 + z T \Bigr \},
\eqn\ee
$$
where $T$ is complex in this representation.
We also expand this with $z$ around $z=0$, and get coefficients $(a,b,c)$
by comparing their coefficients of $z$.
We suppose that $z$ is very small value and we need not $O(z^3)$ terms.

As is stated before, the complex parameter $\alpha$ must be so determined
that the integration will converge. This statement requires that
$Re[(\alpha T)^3]$ is positive for $|T| \gg 1$,
hence there are three $\alpha$'s areas.
In each case, we get the wave function.
The results are
$$
\Psi = Ai(z'')Ai(z'),
\eqn\ee
$$
$$
\Psi = Ai(z'')Bi(z') + Bi(z'')Ai(z') ,
\eqn\ee
$$
$$\eqalign{
\Psi = & Ai(z'') Ai(z') - Bi(z'') Bi(z')                 \cr
       &~~~~~ \pm i  \bigl [ Ai(z'') Bi(z') +Bi(z'') Bi(z') \bigr ],
}\eqn\ee
$$
$$\eqalign{
\Psi = &3 Ai(z'') Ai(z') + Bi(z'') Bi(z')                 \cr
       &~~~~~ \pm i \bigl [ Ai(z'') Bi(z') +Bi(z'') Bi(z') \bigr ],
}\eqn\ee
$$
where the overall factor is neglected.
The wave function (13) has the same form of the Lorentzian
one in Ref.[\HAL]. In this case, the $\alpha$'s area is constrained
on a pure imaginary axis in our method.
Therefore, up to an overall factor,
the path integral form of the (13)'s is the same with the Lorentzian one.
We can show that the wave function
becomes the (13)'s form without using the Wheeler-DeWitt
equation, by following the inverse of the Halliwell's calculation
in Ref.[7] appropriately.
This fact gives us the insight that the singularities in (8)
or (9) is caused for the undesirable integration. We can eliminate
singuralities if we consider the products of Airy functions.
This is why we take $z'=z''=z$.
On the contrary to the (13)'s case, the path integral form of the (14)'s
, (15)'s and (16)'s cases can not be calculated directly.
Therefore we use above insight
and we can calculate the wave function for
other three cases, those are (14), (15) and (16).
In our regularization method, we can follow this insight easily
because we have only turned up the straight contour line
by using the complex parameter $\alpha$.
Of course we must take care of a branch cut caused by
the $ T^{-1/2}$ term in the complex $T$ plane in (12).
We consider both cases that the contour passes through the cut
or not.
{}From this reason, we get six results with three $\alpha$'s areas.
On the other hand, in the contour changing method, the contour is
a complicated
curve. It is very difficult to perform this insight beyond a
saddle point approximation and to distinguish whether the contour
passes through the cut or not.
Therefore the results in Ref.[8] are the part of our results.

Of course we can take another strategy to remove a divergence problem.
If we extend the integrated function as
$$
\Psi (\alpha) =
   \int_{\beta}^{\infty} {d T \over T^{1/2} } e ^{-I(T)}
  +\int^{\beta}_{-\infty} {d T \over (\alpha T)^{1/2} } e ^{-I(\alpha T)},
\eqn\ee
$$
where $\beta$ is some constant, then $\Psi$ also converges.
In this representation, we also fix
$\alpha$ as $Re[(\alpha T)^3]$ is positive for $|T| \gg 1$.
After the integration,
we take the analytic continuation $\alpha \rightarrow 1$.
We also calculate the wave function using
this regularization and the results are almost the same as
the ones by the previous strategy.
However we do not express here.

Beyond a homogeneous and isotropic model, we next analyze
the Bianch type I minisuperspace model \refmark {\HLthr}.
The metric of this type can be expressed as
$$
ds^2={\sigma}^2 \Bigl \{ {N^2(\tau) \over a^2(\tau)} d \tau^2+a^2(\tau)dx^2+
      b^2(\tau) dy^2 +c^2(\tau) dz^2 \Bigr \}.
\eqn\ee
$$
The action of this type is
$$
I(a,b,c) = {1 \over 2} \int_{\tau'}^{\tau''}d \tau
     \Bigl [ -{a \over N}(\dot a \dot b c + a \dot b \dot c+
                          a \dot b \dot c)  +N b c \lambda \Bigr ].
\eqn\ee
$$
In order to deal with the path integral more easily, we change variables of
$a,b,c$ as $x,y,z$ where $x \equiv (bc+a^2)/2$, $y \equiv (bc-a^2)$,
$\dot z^2 \equiv a^2 \dot b \dot c$. The action and the wave function
are
$$\eqalign{
I=&\int_{\tau'}^{\tau''} d \tau (p_x \dot x+p_y \dot y+p_z \dot z-NH),\cr
&H=-p_x^2+p_y^2-{1 \over 2} p_z^2-{ \lambda \over 2} (x+y),
}\eqn\ee
$$
$$\
\Psi=\int dN \int Dp_x Dp_y Dp_z Dx Dy Dz e^{-I}
\equiv \int dN \psi(N).
\eqn\ee
$$
The integral with respect to $x,y,z$ is obtained by means of dividing the
lapse interval $N$ into pieces.
Then $\Psi$ becomes
$$
\Psi=\int {dN \over N^{3/2}} e^{-I_0} ,
\eqn\ee
$$
where
$$\eqalign{
I_0&=-{1 \over 4N} \Bigl \{ (x''-x')^2-(y''-y')^2+2(z''-z')^2
\Bigr \}   \cr
&~~~~~~~~~~~~~~~~~~~~~~~~~~~~~
+{\lambda N \over 4} \Bigl \{ (x'+x'')+(y'+y'')\Bigr \} \cr
&\equiv X N + {Y \over N}.
}\eqn\ee
$$
In this case, the $N$ integration in (22) also diverges,
if we take the integration range of the $N$ from $-\infty$ to $+\infty$.
Therefore
we also use a new regularization method and perform the $N$ integration.
We define again an analytic function by introducing a complex parameter
$\alpha$. After the integration, we take the analytic continuation
$ \alpha \rightarrow 1 $.
On the contrary to a homogeneous and isotropic case, we can calculate
the $N$ integration
explicitly and the result becomes (up to a normalization factor)
$$
\Psi = \sqrt{\pi \over   Y} e^{-2 \sqrt{X Y}}.
\eqn\ee
$$
In Ref.[10], Halliwell and Louko considered the particular Bianch type I model
in which $ b(\tau)=c(\tau)$.
Our result may be a part of the results of the contour changing method.
This reason is that we have treated the integration range of the lapse
function $N$ from $- \infty$ to $+ \infty$
and we have only turned up the straight
contour line again.
By considering a branch cut, we also get a solution that is zero.
However we do not express above.

We finally summarize our results.
In this paper
we have proposed a new regularization method
by concentrating the integration of the lapse function
and evaluated the wave function of the Friedmann-Robertson-Walker model
along this regularization method.
Our results are more general because in our method we can deal with
the branch cut explicitly.
We have then evaluated the wave function of the Bianch type I model.
In this case our result is more limited because we treat the integration
range of the lapse function from $- \infty$ to $+ \infty$.
While we only consider above two simple models, this
regularization method will work well in the more complicated type or
the matter coupled case.

There remain further works with respect to the interpretation
of the wave function by this regularization method.
Here we compare our method to the contour changing method.
One of the most difficult points in the interpretation of the contour
changing method is that its form is neither the Euclidean nor
the Lorentzian path integral \refmark {\Hawwd,\Hawtwo,\HAL,\Gibtwo}.
The wave function following this method can't, in general,
represent the transition
from a purely real Euclidean metric to a purely real Lorentzian one.
This reason is as follows. The contour of this path integral is defined as
the steepest-descent path with respect to the complex lapse function.
Thus the contour is a complicated curve.
In our method on the other hand, by concentrating the lapse function
we remain the real Euclidean form
and evaluate the wave function in a way which is free from
a divergence problem.
The contour of our path integral is defined as the turned up
straight line by using the complex parameter $\alpha$ in an analytic function.
The contour therefore is a simple line.
In our method, we can calculate the wave function of the universe
without using the saddle points and the steepest-decsent path.
We
do not spoil above picture and
can deal with the real tunneling
and get the wave function of the real Euclidean path integral
more easily.

The evaluation of the wave function of the universe is very difficult.
The physical interpretation of the regularization method is
not clear. We consider there exist some possibilities
to obtain a finite wave function along the Euclidean path integral regime,
and our new regularization method is nothing but one of them.

\vskip2.5cm

Acknowledgments

It is a pleasure to thank H. Itoyama to give us many useful
suggestions and comments.
Thanks are also due to S. Sawada for his helpful discussions.
We would like to acknowledge a careful reading of the manuscript by
K. Kikkawa and H. Kunitomo.
We want to express our gratitude to other members of Osaka
University for their warmly encouragements.

\vskip2.5cm

\refout

\end